\documentclass[aps,showpacs ,superscriptaddress,floatfix,twocolumn]{revtex4}
\usepackage{graphicx,} 
\usepackage{amsmath, bbm,amssymb}   
\pdfoutput=1
\begin{document}
\bibliographystyle{prsty}
\title{Qubit decoherence due to detector switching}
\author{I.~Serban}  
\affiliation{IQC and Dept.~of Physics and Astronomy, University of Waterloo, 200 University Ave W, Waterloo, ON, N2L 3G1, Canada}  
\affiliation{Instituut-Lorentz, Universiteit Leiden,P.O. Box 9506, 2300 RA Leiden, The Netherlands}
\author{F.K.~Wilhelm}  
\affiliation{IQC and Dept.~of Physics and Astronomy, University of Waterloo, 200 University Ave W, Waterloo, ON, N2L 3G1, Canada}  
\date{\today}
\begin{abstract}
We provide insight into the qubit measurement process involving a switching type of detector. We study the switching-induced decoherence {\it during} escape events. We present a simple method to obtain analytical results for the qubit dephasing and bit-flip errors, which can be easily adapted to various systems. Within this frame we investigate potential of switching detectors for a fast but only weakly invasive type of detection. We show that the mechanism that leads to strong dephasing, and thus fast measurement, inverts potential bit flip errors due to an intrinsic approximate time reversal symmetry.
\end{abstract}

\maketitle

Noise-activated switching out of a metastable state is a common phenomenon in a wide range of physical systems, including Josephson junctions, nanomechanical devices, chemical reactions \cite{Aldridge05, Wales03}.  Starting with Kramers seminal work \cite{Kramers40}, such processes have been studied close to equilibrium \cite{Hanggi90}, as well as in driven systems \cite{Lehmann00}. The activated escape paths have been studied theoretically and observed experimentally  \cite{Hales08, chan08}. 

Recently, noise-activated switching has gained attention due to its role in quantum measurement, in particular for qubit detection. Examples of switching detectors include the superconducting quantum interference device (SQUID) \cite{Science00,Vion02,Martinis02}, where switching occurs between the superconducting and dissipative state. The Josephson bifurcation amplifier \cite{Siddiqi04,Siddiqi05, Metcalfe07} has been recently employed in the delicate task of detecting a qubit state in a minimally invasive fashion \cite{Lupascu07}.  In this case, the detector can switch between different, weakly dissipative, dynamical states. Using an appropriate choice of a reference frame, switching between such dynamical states can also be described as escape from a {\it static} metastable potential well \cite{Dykman06, MDQT}.

Switching is a highly nonlinear phenomenon, driven by large, non-equlibrium environmental fluctuations, so this type of detection is far from the weak measurement scenario. Some understanding for the switching type of detectors has  been provided by numerical studies \cite{Nakano08,Nakano04} and in a simplified two-state detector version in Ref.~\cite{Ashhab08}. However, a full description of the qubit decoherence during, and induced by the switching event is still missing. 

In this paper we propose a simple and analytical method to investigate qubit decoherence due to a switching type of detector. We model the detector as a classical, overdamped particle trapped in a metastable potential. The escape of the particle is driven by large, rare fluctuations in the environment.  We investigate the qubit dephasing and  bit flip errors induced by the switching, during the escape event. This allows novel insights into the measurement process and reveals the specific conditions during the switching event that lead to a combination of strong coherence loss and low error rate. These are desirable qualities for a qubit detector. 
\begin{figure}[!h]
\includegraphics[width=0.4\textwidth]{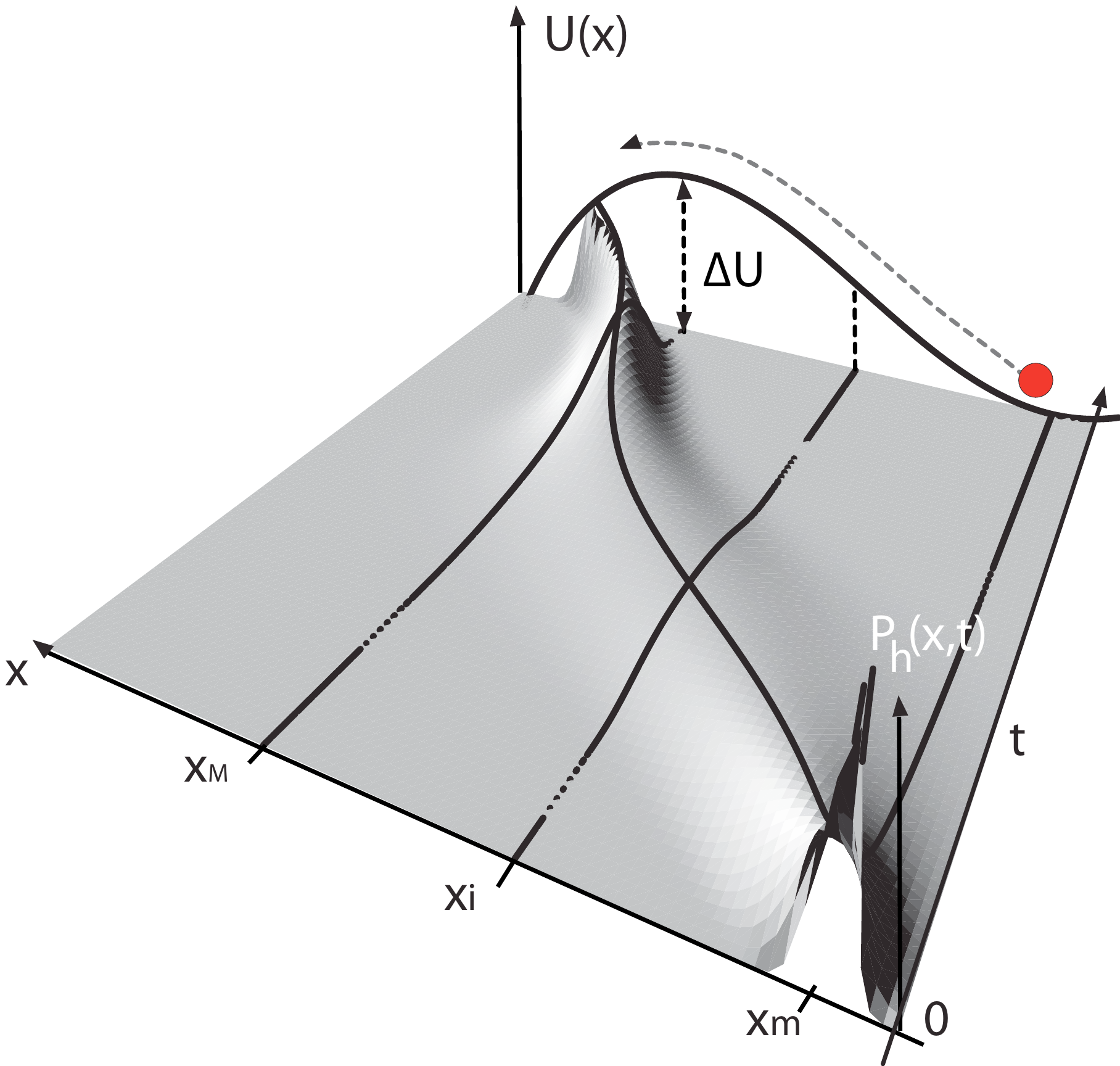}
\caption{Prehistory probability density $P(x,t)$ for metastable potential $U(x)$, where $U'(x)=-K(x)$, and optimal trajectory $x_{\rm opt}(t)$. Here $x_{\rm m,M,i}$  are the positions of the minimum, maximum and inflexion points of the potential.}\label{densill}
\end{figure}
The overdamped classical particle performs Brownian motion according to 
\begin{equation}
\dot x=K(x)+f(t),\label{eq:motion}
\end{equation}
where $K$ is the deterministic force experienced by the particle due to the metastable potential and  $f$ is white Gaussian noise with a probability density functional given by \cite{Feynman65}
\begin{eqnarray}
P[f(t)]&=&\exp\left(-\int_0^{t_{\rm f}}f(t)^2/(2D) dt\right),
\end{eqnarray}
where we assume, \cite{Dykman92}  the noise intensity $D$ to be small compared to the barrier height $\Delta U$, see Fig.~\ref{densill}. The probability density functional  for a noise driven trajectory can be obtained by expressing $f(t)$ in terms of $x(t)$ using Eq.~(\ref{eq:motion}) 
\begin{equation}
P[x(t)]=\exp\left(-\frac{S[x(t)]}{D}\right),\:S[x(t)]=\frac{1}{2}\int_0^{t_{\rm f}}\!\!\!\!\!dt(\dot x-K(x))^2.
\end{equation}

For the study of qubit decoherence during a switching event one will need expectation values of the type
\begin{eqnarray}
O(t_0)&=&\left\langle \exp\left(\lambda\phi[x(t),s(t,t_0)]\right)\right\rangle_{\rm sw},\label{average}\\
\phi[x[(t),s(t,t_0)]&=&\int_0^{t_{\rm f}}\!\!\!\! x(t) s(t,t_0) dt,
\end{eqnarray}
were $s(t,t_0)$ is a time dependent modulation of  $x(t)$. We are interested in the qubit decoherence {\it during} switching. Thus, the averaging $\langle\rangle_{\rm sw}$ is performed  only over switching trajectories of the detector, which satisfy the boundary conditions $x(0)=x_{\rm m}$ and $x(t_{\rm f})=x_{\rm f}$, with $x_{\rm m}$ inside and $x_{\rm f}$ outside the metastable well.  By choosing $s(t,t_0)=0$ at $t_{\rm f}>t>t_0$, the average becomes post-conditioned by a switching event taking place at the final time $t_{\rm f}$. 

Since the exact trajectory between the initial and final point remain unknown, we average over all possible paths 
\begin{eqnarray}
O(t_0)&=&\int_{(x_{\rm m},0)}^{(x_{\rm f},t_{\rm f})}\!\!\!\!\!\!\!\!\!\!\!\!\mathcal{D}x(t)\exp\left(\lambda\phi[x(t),s(t,t_0)]
-\frac{S[x(t)]}{D}\right)\times\nonumber\\
&\times&P(x_{\rm m},0|x_{\rm f},t_{\rm f})^{-1},
\end{eqnarray}
where the total switching probability is
\begin{equation}
P(x_{\rm m},0|x_{\rm f},t_{\rm f})=\int_{(x_{\rm m},0)}^{(x_{\rm f},t_{\rm f})}\!\!\!\!\!\!\!\!\mathcal{D}x(t)\exp\left(-S[x(t)]/D\right).
\end{equation}

The switching  trajectories form a narrow tube in the phase space centered around the optimal trajectory \cite{Onsager53,Dykman89} which minimizes $S$, and for the present case satisfies
\begin{equation} 
\ddot x_{\rm opt}=K(x_{\rm opt})K'(x_{\rm opt}),\: x_{\rm opt}(0)=x_{\rm m},\: x_{\rm opt}(t_{\rm f})=x_{\rm f}.\label{motion1}
\end{equation} 
Thus $S[x(t)]=S[x_{\rm opt}(t)]+S_2[x(t)-x_{\rm opt}(t)]$
and we approximate
\begin{equation}
S_2[x(t)]\approx\frac{1}{2}\int_0^{t_{\rm f}} dt (\dot{x}(t)^2-\Lambda(t)^2 x(t)^2), \label{semiclassics}
\end{equation}
where $\Lambda(t)^2=-(K'(x)^2+K(x)K''(x))|_{x=x_{\rm opt}(t)}$. Divergences due to the emergence of a slow mode on the barrier top \cite{Dykman92, Dykman98} are avoided by the appropriate  choice of the initial (kinetic) energy $0<\dot{x}(0)^2/2
\ll\Delta U$ which satisfies the boundary conditions (\ref{motion1}). Thus, the switching event takes place, with non-vanishing probability, within a {\it finite} time $t_{\rm f}$.
One can show that
\begin{equation}
O(t_0)=\exp\left(\lambda\phi[x_{\rm opt}(t)+x_0(t)/2,s(t,t_0)]\right),\label{result}
\end{equation}
where $x_0$ is the solution of
\begin{equation}
\ddot x_0+\Lambda^2 x_0+D\lambda s(t,t_0)=0,\: x_0(0)=x_0(t_{\rm f})=0.\label{motion2}
\end{equation}
The two linearly independent solutions of the homogeneous part of Eq.~(\ref{motion2}) are
\begin{equation}
x_1(t)=\dot x_{\rm opt}(t),\: x_2(t)=\dot x_{\rm opt}(t)\int_0^t dt'\dot x_{\rm opt}(t')^{-2},
\end{equation}
and the full $x_0(t)=x_1(t)c_1(t)+x_2(t)c_2(t)$ can be determined by variation of constants.

We consider the case of a metastable potential described by $U(x)/\Omega=x/2-x^3/6$, which can represent a Josephson junction DC-biased at half the critical current. The characteristic frequency of the detector is given by $\Omega=K'(x_{\rm m})$.

{\it Pure dephasing}: 
We assume a qubit Hamiltonian of the form
\begin{eqnarray}
\hat{H}&=&\hbar\omega\hat\sigma_z+\eta\hat{\sigma}_z x(t),
\end{eqnarray}
where $x(t)$ is the coordinate of the classical particle. The only effect of the environment in this case is the irreversible decay of the phase coherence $C(t_0)=O(t_0)$, see Eq.~(\ref{average}), for the specific value of $\lambda=\eta/(i\hbar)$ and 
\begin{eqnarray}
s(t,t_0)&=&\left\{\begin{matrix}
1,&t<t_0\\
0,&t>t_0.
\end{matrix}\right.\label{s_choice}
\end{eqnarray}

\begin{figure}[!h]
\includegraphics[width=0.35\textwidth]{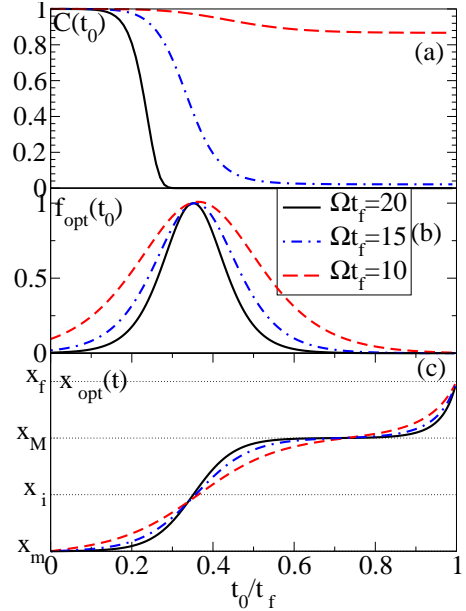}
\caption{Qubit coherence during a switching event $(t_0<t_{\rm f})$ (a), optimal noise trajectory $f_{\rm opt}(t_0)$ (b) and most probable switching trajectory $x_{\rm opt}(t_0)$ (c)  for various values of $t_{\rm f}$.}\label{coherence}
\end{figure}

Fig.~\ref{coherence} shows that the escape process is driven by strong noise (b), with a maximum intensity at the time where the optimal trajectory (c) reaches the inflexion point of the barrier. In the vicinity of the same point, a sharp drop in qubit coherence (a) is observed. After the inflexion point the motion is slowed down, and becomes diffusive close to the barrier top. During this stage, the qubit coherence remains at an almost constant value. We observe that the optimal noise becomes stronger for shorter values of $t_{\rm f}$. However, the strongest drop in qubit coherence was observed for the longer $t_{\rm f}$. In this case the optimal trajectory spends more time close to the barrier top, where the motion is diffusive, driven by low amplitude noise.

{\it Bit flip errors}:  We consider a qubit-environment coupling which allows for energy exchange, and can induce bit flip errors
\begin{eqnarray}
\hat{H}=\hbar\omega\hat\sigma_z+\eta x(t)\hat \sigma_x.
\end{eqnarray}
The probability of noise induced errors during a switching event, at $t_0<t_{\rm f}$ is given by
\begin{equation}
P_{\uparrow\to\downarrow}(t_0)=|\langle\downarrow|\hat{U}_I(t_0)|\uparrow\rangle|^2\: P(x_{\rm m},0|x_{\rm f}, t_{\rm f})^{-1},
\end{equation}
where in the limit of short time and weak coupling
\begin{eqnarray}
\hat{U}_I(t_0)&=&\mathcal{T}\exp\left(\int_0^{t_0}\!\!\!\! dt \frac{H_I(t)}{i\hbar}\right)\approx 1+\int_0^{t_0}dt\frac{H_I(t)}{i\hbar},\nonumber\\
\hat{H}_I(t)&=&\eta\hat{U}_0^\dagger(t)\hat \sigma_x\hat{U}_0(t)x(t),
\end{eqnarray}
and $\hat U_0$ describes the free qubit evolution. We obtain
\begin{eqnarray}
P_{\uparrow\to\downarrow}(t_0)&=&\lim_{\lambda\to0}\frac{\partial^2}{\partial \lambda^2}\frac{\eta^2}{\hbar^2}\left\langle
\exp\left(\lambda\phi[x(t),s(t,t_0)\mathcal{R}(t)]\right)\right.\nonumber\\
&+&\left.\exp\left(\lambda\phi[x(t),s(t,t_0)\mathcal{I}(t)]\right)\right\rangle_{\rm sw},\nonumber\\
\mathcal{R}(t)+i\mathcal{I}(t)&=&\langle\downarrow|\hat{U}_0^\dagger(t)\hat \sigma_x\hat{U}_0(t)|\uparrow\rangle.
\end{eqnarray}

\begin{figure}
\includegraphics[width=0.35\textwidth]{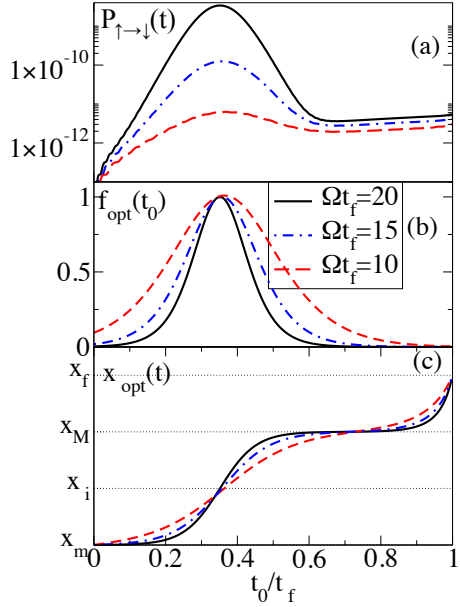}
\caption{Probability for an induced bit flip error as a function of $t_0$, for various values of $t_{\rm f}$ (a), the corresponding most probable noise (b) and the most probable driven trajectory (c). Here $\omega=10 \Omega$. }\label{relaxation_fig}
\end{figure}

In Fig.~\ref{relaxation_fig} (a) we observe, similar to the pure dephasing case, a sharp feature in $P_{\uparrow\to\downarrow}(t_0)$ at the point in time where the most probable trajectory (c) reaches the steepest point on the potential barrier, and the most probable noise (b) reaches it maximum strength. Despite the optimal noise being strongest for short switching time $t_{\rm f}$, the peak in $P_{\uparrow\to\downarrow}(t_0)$ is higher for longer $t_{\rm f}$. Another notable feature is the quasi-reversibility of the bit flip error which occurs at this point. This feature cannot be explained by the single, deterministic trajectory $x_{\rm opt}$ alone. It causes only the steady increase of $P_{\uparrow\to\downarrow}(t_0)$.

{\it Prehistory density distribution}: The results presented above can be understood from the distribution of switching trajectories. We calculate the probability $P_h(x,t)$ for the classical particle to occupy the position $x$ at time $t$ during a switching event,  in the form of a prehistory density distribution \cite{Dykman92}
\begin{eqnarray}
P_h(x,t)=\frac{P(x_{\rm m},0|x,t)P(x,t|x_{\rm f},t_{\rm f})}{P(x_{\rm m},0|x_{\rm f},t_{\rm f})}. 
\end{eqnarray}
Within the approximation (\ref{semiclassics}), the probability for a transition between any pair of points $(x_1,t_1)$ and $(x_2, t_2)$, with $t_{1,2}<t_{\rm f}$ reads
\begin{equation}
P(x_1,t_1|x_2,t_2)=\int _{(\delta x_1,t_1)}^{(\delta x_1,t_2)}\!\!\!\!\!\!\!\!\!\!\!\!\!\!\!\!\mathcal{D} x(t)\exp\left(-\frac{S[x_{\rm opt}(t)]_1^2+S_2[x(t)]_1^2}{D}\right),
\end{equation}
where $S[x(t)]_1^2$ implies that the time integral is taken between $t_1$ and $t_2$ and $\delta x_{1,2}=x_{1,2}-x_{\rm opt}(t_{1,2})$. One can show that
\begin{eqnarray}
P(x_1,t_1|x_2,t_2)&=&\exp\left(-\frac{S[x_{\rm opt}(t)]_1^2+S_2[x_{b}(t)]_1^2}{D}\right)\nonumber\\
&\cdot&F(t_1|t_2),\label{prehistory}
\end{eqnarray}
where $\ddot{x}_b+\Lambda^2(t)x_b(t)=0$ and $x_b(t_{1,2})=\delta x_{1,2}$ and
\begin{eqnarray}
F(t_1|t_2)&=&\int_{(0,t_1)}^{(0,t_2)}\!\!\!\!\mathcal{D} x(t)\exp\left(-\frac{S_2[x(t)]}{D}\right)\\
&=&\left(2\pi D \dot x_{\rm opt}(t_1)\dot x_{\rm opt}(t_2)\int_{t_1}^{t_2}\dot{x}_{\rm opt}(t)^{-2}dt\right)^{-1/2}. \nonumber
\end{eqnarray}
We obtain a Gaussian distribution, centered around $x_{\rm opt}(t)$
\begin{eqnarray}
P_h(x,t)=\frac{1}{\sqrt{\pi w(t)}}\exp\left(-\frac{(x-x_{\rm opt}(t))^2}{w(t)}\right),
\end{eqnarray}
where $w(t)=F(0|t_{\rm f})^{2}/(F(0|t)^2F(t|t_{\rm f})^2)$.
\begin{figure}[!h]
\includegraphics[width=0.45\textwidth]{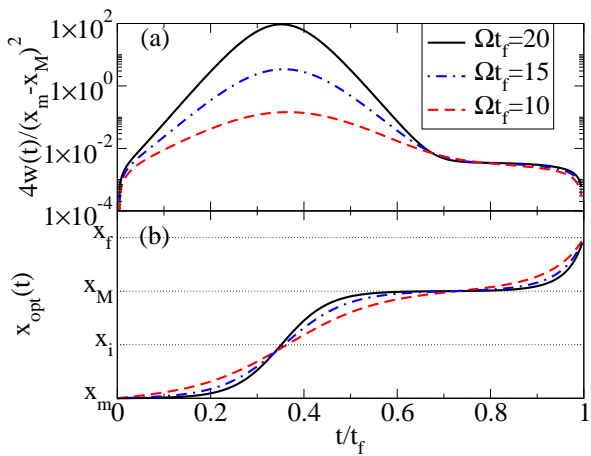}
\caption{Width of the prehistory probability distribution $P_h(x,t)$ (a) and optimal trajectory (b).}\label{dens}
\end{figure}

Fig.~\ref{dens} shows a narrow tube of trajectories close to the bottom of the well. This is followed by a strong widening of the distribution in the process of climbing up the potential barrier. This event is driven by a sharp noise pulse.  On the barrier top we see again a fairly localized density distribution, driven by low-amplitude noise. The tube narrows even more on the outer side of the barrier. These results are in agreement with the findings of Ref.~\cite{Dykman00}, for a different system.

We found, see Figs.~\ref{coherence} and \ref{relaxation_fig} that the qubit suffers the strongest decoherence at the point in time when the optimal trajectory reaches the steepest point on the barrier wall. This is true for both bit flip errors and dephasing. The magnitude of both effects depends strongly on the total time necessary for the switching event, such that longer $t_{\rm f}$ leads to enhanced coherence loss, and higher bit flip rate. The observed effect can be explained by the strong widening of the prehistory distribution $P_h(x,t)$ at the same point in time, see Fig.~\ref{dens}.

The peak in $P_{\uparrow\to\downarrow}$ originates in the strong widening of the trajectory tube. Thus, large excursions around $x_{\rm opt}$, see Fig.~\ref{densill}, are very probable during this time. However, since each of these switching trajectories must return to the narrow tube of trajectories on the other side of the widening, i.e. in the region close to the barrier top, they all show an approximate time reversal symmetry. Thus any induced bit flip errors will be quasi reversed when the particle reaches the barrier top.

In conclusion, a switching detector, such as the one modeled here, presents several qualities which make it desirable as a nonlinear qubit detector. The strong decoherence suffered by the qubit as the classical particle climbs up the potential barrier affects strongly the coherence, leading to a fast measurement. The equally strong bit flip errors acquired during the process are reversed by the quasi time-reversal symmetry of most trajectories.  

We observe that a fast switching event causes less decoherence, despite stronger noise being required for surmounting the barrier. Fig.~\ref{dens} reveals that longer switching time $t_{\rm f}$ allows  more freedom in the choice of the particular time the particle climbs up the potential barrier and such incoherent behavior causes decoherence.
 
We note that the widening of the trajectory tube at the inflexion point of the potential appears throughout literature as a general feature of the tube of escape trajectories out of metastable potentials. Having identified it as the major cause for the features observed in the qubit decoherence, we expect these features to be common to various potentials.  Therefore we expect that our results have applicability to existing experimental setups, e.g. the JBA and the DC-SQUID.  

We are grateful to M.I. Dykman for pointing out the prehistory probability distribution approach and for many useful suggestions.
We acknowledge useful discussions with J. Gambetta, W.A. Coish and T.C. Wei. This work has been supported by NSERC Discovery Grants and Quantum Works.

 \end{document}